\documentstyle[11pt,paspconf,epsfig]{article}

\begin{document}

\title{Super-Eddington accretion in GRS 1915+105 }

\author{O. Vilhu}
\affil{Observatory, Box 14, FIN-00014 University of Helsinki, Finland}

\begin{abstract}

Classical modelling suggests that, during the
RXTE observations studied, GRS 1915+105 was near or above  
the critical accretion rate and the optically thick inner disk 
penetrated inside the
advection-dominated flow. 
The system was very unstable leading
to a rich pattern of variability, e.g. the {\it rings}
which might
be characteristic to disk instabilities very close to the innermost stable
orbit.
Small values of R$_{in}$ obtained  require a rapidly rotating Kerr 
hole, unless the central mass is smaller than 2-3 M$_{\sun}$. 
A thermal sombrero provides a reasonable modelling of the rings, although
the broad-band RXTE/OSSE spectrum
analysed  seems to require a significant injection of relativistic
electrons, the hot phase being not a pure thermal one.
Observations below 2 keV and simultaneous
broad-band spectra between 2--400 keV are needed to better fix the 
size of the 
black body disk and the role of relativistic injection. 
Further,
theoretical work on emerging spectra of Kerr-holes 
 is required to replace the classical model used.

\end{abstract}


\section{Introduction}
The Galactic X-ray transient GRS 1915+105 was discovered by Castro-Tirado,
Brandt \& Lund (1992). After Mirabel and Rodriguez (1994) found superluminal
jets in this black hole candidate,
it received the nickname `microquasar'. 
A rich pattern of X-ray variability has
emerged from the Rossi X-ray Timing Explorer (RXTE) showing a great potential
to study accretion phenomena close to a black hole (e.g. Greiner et al. 1996,
Morgan et al. 1997, Chen et al. 1997, Taam et al. 1997, Belloni et al. 1997a,b, 
Markwardt, 1997, Vilhu and Nevalainen 1998).

In the present paper I discuss important parameters of the accretion
theory, the accretion rate and the inner disk radius, as derived from 
classical disk models and RXTE observations. 
Two-phase thermal modelling of the {\it rings}
discovered by Vilhu and Nevalainen (1998) is also presented.

\section{ Accretion rate and  inner disk radius  }

\begin{figure}
\begin{center} \leavevmode
\hbox{%
\epsfysize=7cm \epsffile{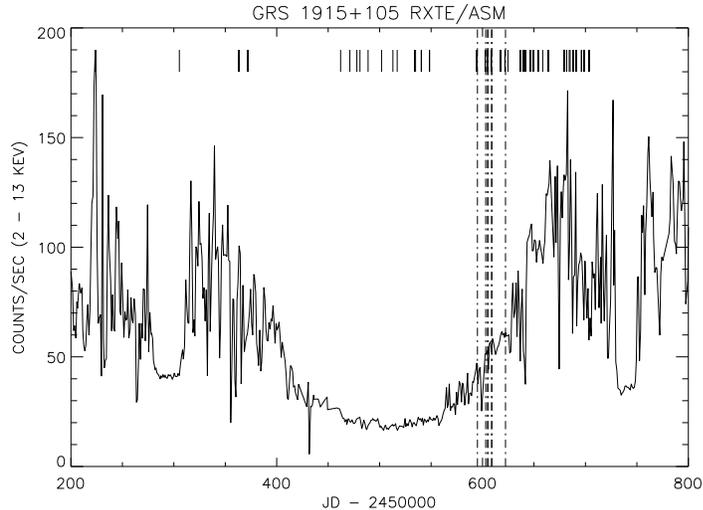}}
\end{center}
\caption{                
The RXTE/ASM light curve (2--13 keV) of GRS 1915+105.
The ring-observations are marked with vertical dashed lines.
The short lines show those PCA observations used to construct Fig.2.} 
\label{fig-1}
\end{figure}

Several RXTE/PCA observations (see the vertical lines in 
Figure~\ref{fig-1}) were analysed 
 using the {\it diskbb + power law} model of the 
XSPEC software (Vilhu and Nikula, in prep.) . The disk model assumes that the viscously released energy
is radiated away locally, without any advection (see Frank et al. 1992). 
Temperature scales with radius as T = T$_{in}$(r/r$_{in}$)$^{-3/4}$ and
the mass accretion rate  is given by     
$\dot M$=8$\pi$R$_{*}$$^3$$\sigma$T$_{*}$$^4$/3GM where T$_{in}$ = 
0.5T$_{*}$ 
and r$_{in}$ = 49r$_{*}$/36. This modelling is consistent with the disk      
luminosity L$_{disk}$ = $\eta$$\dot M$c$^2$ {\bf if} the radiation efficiency
is computed from $\eta$  = 0.12(r$_{in}$/r$_g$)$^{-1}$  where r$_g$ = 2GM/c$^2$ is 
the gravitational radius. (At 3r$_g$ this gives an efficiency 0.04 which is
not far from 0.057 estimated by Zycki et al. (1997).) 
Further, we define the Eddington luminosity  and accretion rate as 
L$_E$ = 4$\pi$GMm$_p$c/$\sigma$$_T$ and $\dot M$$_E$ = L$_E$/c$^2$,
respectively.
 
\begin{figure}
\begin{center} \leavevmode
\hbox{%
\epsfysize=7cm \epsffile{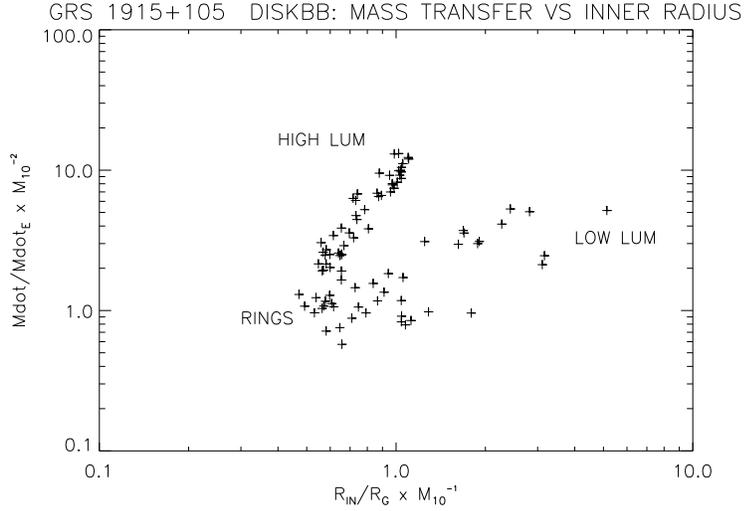}}
\end{center}
\caption{                
Results from classical disk modelling of observations marked in
Fig.1. Along the boomerang-shaped pattern (from LOW LUM
to HIGH LUM) the disk luminosity, the PL-luminosity,
the inner disk temperature 
and the photon index vary between  
0.1 - 2.5 L$_E$, 0.2 - 0.6 L$_E$, 0.5 - 2.5 keV and 
2.2 - 3.5, respectively ( 10M$_{\sun}$) .}
\label{fig-2}
\end{figure}

\begin{figure}
\begin{center} \leavevmode
\hbox{%
\epsfysize=7cm \epsffile{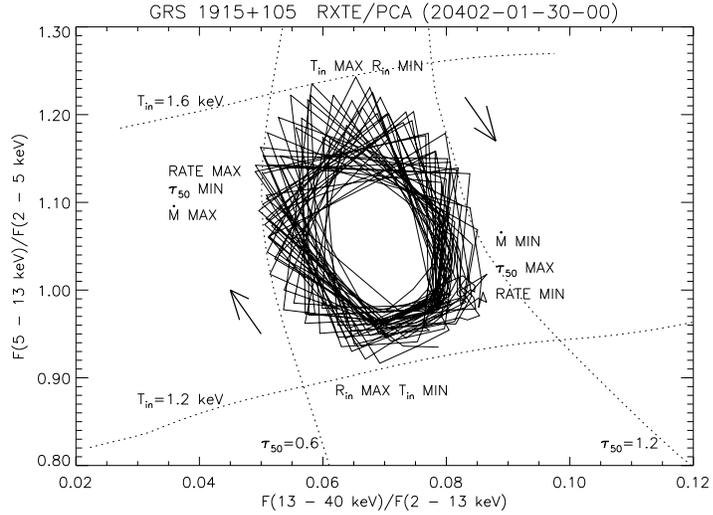}}
\end{center}
\caption{                
A ring-observation 
  plotted in the 2-colour diagram with 16 sec time-binning
(from Vilhu and Nevalainen, 1998). 
The big arrows show
the clock-wise evolution with a 97 s mean cycle period.    
 2-phase model curves (thermal sombrero) 
are overplotted with dotted lines.} 
\label{fig-3}
\end{figure}

\begin{figure}
\begin{center} \leavevmode
\hbox{%
\epsfysize=7cm \epsffile{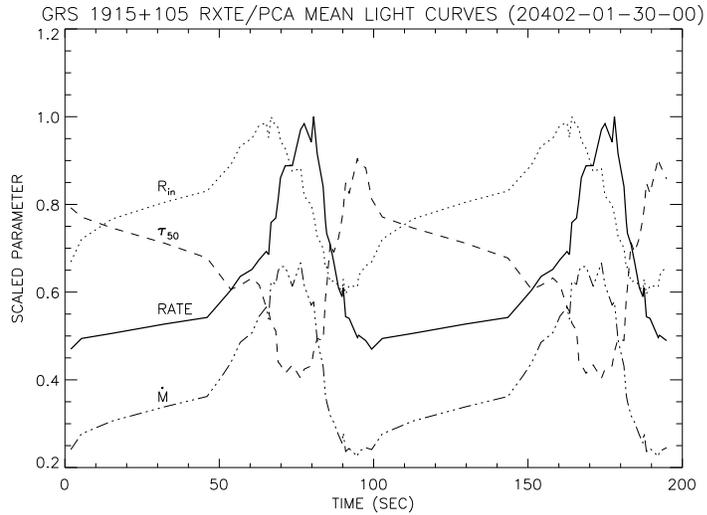}}
\end{center}
\caption{                
Mean light curves over one cycle for the observation
of Figure~\ref{fig-3}. 
The parameters are scaled by the following values (in parentheses):
RATE (15000 cts/s (5 PCU units), solid line), R$_{in}$ (30 km, dotted line),
$\tau$$_{50}$ = $\tau$(T$_e$/50) (1.4, dashed line) and 
$\dot M$ (3L$_{E}$/c$^2$ of 10M$_{\sun}$, dash-dotted line). }
\label{fig-4}
\end{figure}

\begin{figure}
\centerline{\epsfig{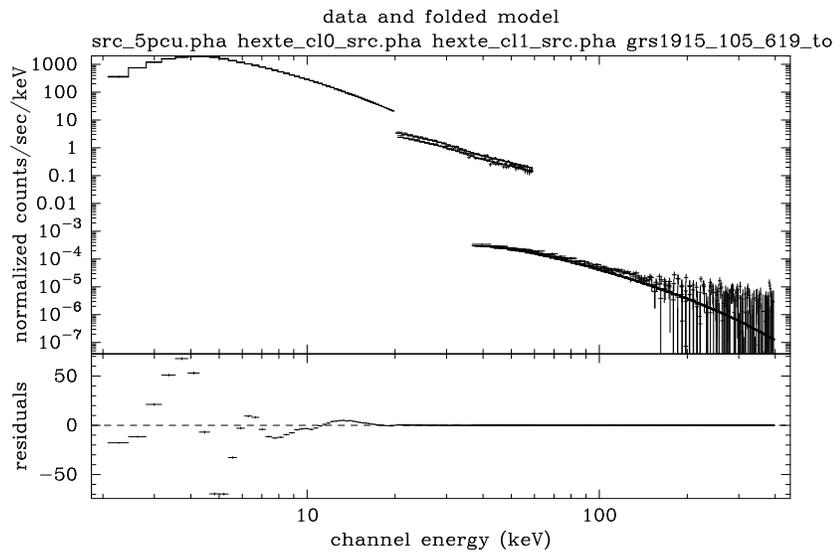}}
\caption{An average ring spectrum  
(RXTE/PCA/HEXTE and CGRO/OSSE)  with a thermal sombrero fit. 
The high residuals between 
4-5 keV are due to the istrumental artefact caused by the Xe 
L-edge of the PCA. The high energy tail clearly requires an additional 
injection of 
non-thermal relativistic electrons.}
\label{fig-5}
\end{figure}

The results of this classical modelling are shown in Figure~\ref{fig-2}
(r$_{in}$ in units of r$_g$ and $\dot M$ in units of M$_E$ of 10M$_{\sun}$).
Due to advection, the  accretion rates  are probably 
 lower limits. Further studies of realistic disk models and their
'skin' spectra should  confirm (or disprove) the small inner disk radii
derived from this classical treatment (being as small as 15 km).
First attempts in this direction have already been taken by e.g.
Beloborodov (1998). However, the present results indicate 
 that the inner disk reaches 0.5r$_g$ if the mass
is  10M$_{\sun}$ and 3r$_g$ in the case of 1.5M$_{\sun}$.
In the more massive case we clearly need a rotating Kerr-hole which
would not be a surprise due to the existence of jets. (Note that the  
innermost stable orbits are at 3r$_g$ and 0.5r$_g$ for non-rotating
Schwarzschild and extreme Kerr holes, respectively.)

In the 10M$_{\sun}$ case the inner disk lies  
 inside the trapping radius where
 a significant part of the energy is advected to the black hole instead 
of being radiated away (see Beloborodov, 1998).  The only solution
allowed is the advection-dominated flow (see Chen et al. 1995).
How it can also contain
an optically thick component (as indicated by the results presented here) 
remains to be studied. However, the solution must be very unstable due to the
high variability  seen in all time scales from months to seconds. In the next
Section we discuss one special type of variability.

\section{ Rings                  }

Vilhu and Nevalainen (1998) found
a peculiar X-ray variability in a narrow range of luminosity 
and called them {\it rings} 
(see the dashed lines in Figure~\ref{fig-1}).
In Figure~\ref{fig-2} the rings are situated at the knee of the V-type 
pattern (at 
smallest accretion rates and smallest inner disk radii).
This variability was found as a ring-shaped pattern 
in the two-colour diagram of count
rates, where the
hard hardness F(13-40keV)/F(2-13keV) is plotted against the soft hardness
F(5-13keV)/F(2-5keV). The system runs one cycle with  periods
ranging between 50--100 s for different observations, one rotation 
in the 2-colour diagram corresponding 
to the time between two contiguous maxima in the light curve.     

An example of such a behaviour is shown in Figures 3 and 4 and modelled 
with a thermal sombrero-model (see Poutanen, 1998),
where the optically thick classical disk 
penetrates into the spherical hot corona (ISMBB in our XSPEC). 
This geometry is 
similar to that of the popular advection-dominated accretion flow 
(see e.g. Esin et al 1997). 
The model is basically the same as in the previous Section, except 
that the power-law is  physically explained by Comptonization of soft
disk photons in the central spherical hot phase.
During  these particular observations 
the inner disk radius  varied with a 97 s period 
between 20--35 km with an anticorrelation
between the coronal $\tau$T$_e$ and the mass accretion rate $\dot M$,
possibly indicating a coupling between the disk and coronal accretion.

To check the thermal/non-thermal nature of the radiation, one has to look
at the hard part of the spectrum.
Figure~\ref{fig-5} shows such an
average ring-spectrum which occasionally had nearly simultaneous OSSE/CGRO 
observations available (from Vilhu et al. 1998). A moderate fit with 
a thermal sombrero gave:
$N_H=2.27\cdot10^{22}$ cm$^{-2}$, $T_{in}$=1.57 keV, $T_{cor}$=68 keV and 
$\tau_{cor}$=0.67, compatible with those found from the model curves 
of Figure~\ref{fig-3}. It can be seen that the thermal model alone does not 
reproduce well the hard part of the spectrum.
Due to this hard tail, the electron distribution
probably deviates from the Maxwellian one. A fraction of the energy 
input could be injected in the form of relativistic electrons (pairs). 
In the  EQPAIR-model (again in the XSPEC software)  
this is allowed in addition to the direct heating of 
thermal electrons (Coppi et al 1998). Using  this model for
the observations of Figure~\ref{fig-5},
the non-thermal efficiency becomes 28 \% ($l_{nth}/l_h=0.28$)
while the ratio of the soft and hard compactnesses  $l_h/l_s$ equals  0.36. 
Most importantly, the fit is  much better 
than in the case of the thermal sombrero.

\section{Discussion}                    

Using classical modelling for the RXTE data suggests that, during the
observations studied, GRS 1915+105 was above or 
near the critical accretion rate where the accretion luminosity exceeds the 
Eddington one. The optically thick inner disk penetrated inside the
advection-dominated inner trapping flow. 
The system is very unstable leading
to a rich pattern of variability, e.g. the {\it rings}.
 
 The ring-behaviour might
be characteristic to disk instabilities very close to the innermost stable
orbit,     where the relativistic potential barrier is small 
and just 
a small change in angular momentum leads to accretion through the horizon.
 However, 
the minimum values of r$_{in}$ obtained (20 km) are too small for a 
non-rotating black hole and 
require a rapidly rotating one,
unless the central mass is smaller than 2-3 M$_{\sun}$. 
Rapid rotation of the hole might be the energy reservoir for
the blobs (jets) ejected frequently at almost the velocity of light.

Broad band X/$\gamma$ ray spectra of black hole candidates have revealed that
the radiation processes can be explained in terms of a disk black body and
successive Compton scatterings of soft photons (Comptonization) in a hot 
electron cloud surrounding the object.  However, the broad-band spectrum
analysed  seems to require a significant injection of relativistic
electrons, so the hot phase is not a pure thermal one.

One can safely conclude that observations below 2 keV and more simultaneous
broad-band spectra between 2--400 keV are needed to better fix the 
size of the 
inner disk and the role of relativistic injection. Future satellites, such as
XMM, SRG and INTEGRAL, are well tailored for these kinds of observations. 
In addition,
theoretical work on `skin'-spectra of Kerr-holes 
are necessarily required to replace the classical models used in the
present paper.




\end{document}